\documentclass[10pt]{scrartcl}

\title{Tagging and Linking Lecture Audio Recordings: Goals and
  Practice\thanks{Some of the
  material here was presented in a conference paper at
  \emph{International Enhancement Themes: Enhancement and Innovation}, Glasgow, June 2013}}
\author{Norman Gray${}^1$\thanks{Corresponding author: \texttt{norman@astro.gla.ac.uk}},
    Nicolas Labrosse${}^1$,\\
    Sarah Honeychurch${}^2$, Steve Draper${}^3$, Niall Barr${}^2$\\
    $^1$ School of Physics and Astronomy,\\
    $^2$ Learning and Teaching Centre, and\\
    $^3$ School of Psychology, University of Glasgow, Glasgow, UK}
\date{7 November 2013}

\usepackage{url,graphicx}

\usepackage{prettyref}
\newrefformat{s}{Sect.~\ref{#1}}
\newrefformat{f}{Fig.~\ref{#1}}

\let\paraheader\emph

\usepackage{hyperref}

\begin{document}
\maketitle


\begin{abstract}
Making and distributing audio recordings of lectures is cheap and
technically straightforward, and these recordings represent an
underexploited teaching resource.  We explore the reasons why such
recordings are not more used; we believe the barriers inhibiting such
use should be easily overcome. Students can listen to a lecture they
missed, or re-listen to a lecture at revision time, but their
interaction is limited by the affordances of the replaying technology.
Listening to lecture audio is generally solitary, linear, and disjoint
from other available media.

In this paper, we describe a tool we are developing at the University
of Glasgow, which enriches students' interactions with lecture audio.
We describe our experiments with this tool in session 2012--13. Fewer
students used the tool than we expected would naturally do so, and we
discuss some possible explanations for this.
\end{abstract}

\begin{centering}
\bfseries
*** This is an early DRAFT for discussion ***;\\
a subsequent version is intended for submission to\dots TBC.\\
\end{centering}

\section{Introduction}

Making audio recordings of lectures is cheap (in money and time), and
technically straightforward.  Together, these mean that it is easy for
lecturing staff to create this additional resource without much in the
way of support, which in turn makes it easy for them to do so
routinely and robustly, with little intellectual or technical buy-in.
It is also reasonably easy to distribute the audio to students, and
people have in the past done so using VLEs or services such as Apple's
iTunes.

It is hard to escape the feeling, however, that while it is easy to
make recordings, they are hard to exploit fully: there is more value
in lecture recordings than is readily accessible.  Students can listen
to a lecture they missed, or re-listen to a lecture at revision time,
but their interaction is limited by the affordances of the replaying
technology.  Listening to lecture audio is generally solitary, linear,
and disjoint from other available media.

In this paper, we describe a tool we are developing at the University
of Glasgow, which enriches students' interactions with lecture audio.
We describe our experiments with this tool in session 2012--13.

Our general ambitions are:
\begin{itemize}
\item to elicit (and share) student generated content in the form of
  tags attached to audio instants, and links between the audio and
  other lecturer- or student-generated material;
\item to enable and encourage students to interact with the available
  material, which helps them reprocess it intellectually through,
  amongst other things, a type of prompt rehearsal;
\item to support that reprocessing with pedagogically well-founded exercises and activities; and
\item to enable (`empower') students to interact with institutionally
  provided materials, on multiple devices (including mobile), in an
  attractive and up-to-the-minute style.
\end{itemize}
In practice, the `audiotag' tool organises and distributes lecture
recordings, supports tagging instants within the audio, and supports
peer `likes' of those tags; see \prettyref{s:audiotag} for more detail.

During session 2012--13, the Audiotag team received funding from
Glasgow University (i) to formally evaluate the audiotag service in
the context of lecture courses across the university, (ii) to evolve
it towards greater usability, (iii) to develop teaching techniques to
help students exploit the service possibilities, and (iv) to work with
a student developer revisiting the interface and imaginatively
exploiting the available service ecology, with cross-links to other
media.

To our surprise, we report below a suprisingly low engagement with the
audio lectures, on the part of the students we have worked with, which
has frustrated our attempts to devise more interesting pedagogical
exercises.  We discuss some possible explanations for this.

In \prettyref{s:background} we describe some of the motivating
background for our current work.  In \prettyref{s:audiotag} we
describe the software system we have developed to support this work,
and in \prettyref{s:experiment} the results of using this tool to support
a set of six lecture courses in astronomy.  Finally, in
\prettyref{s:summary} we reflect on the results we have obtained.

\section{Background and motivation}
\label{s:background}
It is still relatively uncommon for lecturers to make available
recordings of their lectures.  The latest Digital Natives
survey~\cite{gardiner11} shows that 90\% of students `expect' lecture
recordings, so there is at least some, possibly somewhat unfocused,
demand for them. Basic audio-recordings of lectures are easy to
produce and distribute (creating a podcast is both cost- and
time-efficient) so that there are few real technical or cost barriers
to making recordings available. Though there is often some scepticism
about the practice, in our experience relatively few lecturers are too
shy to have their words recorded, or raise for example intellectual
property concerns.  Why, then, is lecture recording not ubiquitous?

We can find some explanation by looking more closely at the supply of
recordings, the demand for them, and the pedagogical justification for
and use of them; we find something resembling a vicious circle.
We believe that the supply barriers are deemed
significant because the demand is too low, the demand is low (or at
least too vague) because the student body is unfamiliar with the
possibility and so does not know to ask for a supply, and the
pedagogical benefits (which might cause lecturers to create the supply
irrespective of demand) are underexplored because too few lecturers
use the technique for them to successfully explore the space of
possibilities.

\paraheader{Supply:} Digital voice recorders are now inexpensive or ubiquitous (they range
from \pounds 30--\pounds150, and many smartphones have adequate
recording capabilities out of the box), most people seem to have reasonably
ready access to basic audio-editing software, and they can distribute
audio files by uploading them to the university Moodle servers.
Several of the current group used the free application `Audacity' to
make minimal edits\footnote{See
  \url{http://audacity.sourceforge.net}}, which took perhaps 15
minutes of effort after a lecture; we do not expect lecturers (or
support staff) to do any elaborate post-production beyond, perhaps,
top-and-tailing or de-noising, and in particular we do not expect
anyone to produce anything more sophisticated than a reasonably
audible hour of one individual's monologue. The final step of making a
podcast from the audio collection is more intricate, but Moodle, like
many similar services, has a podcasting plugin\footnote{The
  distinction between a podcast and a mere collection of audio files
  is the presence of a `feed' -- an RSS or Atom file -- which allows a
  `feed reader' application to be automatically notified of the
  appearance of new `episodes', so that a user doesn't have to
  repeatedly re-check the audio source.}.  Each of these technical
obstacles is by itself relatively minor, but in combination they are a
barrier substantial enough that only an enthusiast would currently
breast them.

There is also a type of `supply' question from the students' side, in
the supply of technical expertise which students can already be
assumed to possess.  Students (or the younger ones at least) have been
described as `digital natives', more than 98\% of whom have ready
access to a computer, 65\% of whom share photos on social networks,
and 20\% of whom even report that they edit audio or video, at some
level, on a monthly basis.  Given this, it is very tempting to assume
that there is little or no effective barrier to students' uptake of
reasonably straightforward learning technology.

\paraheader{Demand:} It is not particularly surprising that a large
fraction of students report -- in both formal and informal
feedback -- that they would welcome lecture
recordings~\cite{gardiner11}.  However this does not appear to be reflected
in actual usage figures when the recordings are made available (see
also the usage analysis below in \prettyref{s:results}).

One likely reason for this is that an hour-long recording is not a
particularly usable format: it may be useful to provide a
`listen-again' opportunity on a long commute, but the devices that
students naturally use to listen to podcasts, being primarily targeted
at either music or at podcasts patterned after magazine-style radio
programmes, are not easy to use for dipping into, or referring to
chunks within, a long recording.

We discussed the usage of recordings with small groups of students
shortly after the corresponding exams, so after the students had had
the opportunity to investigate or reinvestigate their potential as a
revision aid.  The students stressed that the recordings were
in practice notably more useful for some material than for others.  A
course, or a section, which was ``just maths and facts'' (to quote one
student) might be more effectively revised using printed notes or slides,
rather than a recorded oral explanation; in contrast, the recording
might be the most valuable resource for more conceptual material.
Whatever this implies about learning modalities and strategies,
it is clear that it adds an extra variable to what material we should
expect to be useful in a particular context.

\paraheader{Pedagogic utility:} Despite the lack of an urgent demand from
our intended users, we believe that there is a great deal of
educational value latent within lecture audio.  This arises partly
from its pragmatic use as a revision aid, but also, more
fundamentally, because it represents a different modality for
instruction, which may complement or in extreme cases replace more
traditional textual routes for some students.   From this position is
it natural to investigate that use of our system within a
peer-assisted learning technique such as Jigsaw~\cite{aronson13},
which members of our team have already successful used within the
university; in the event, however, we have not yet had the opportunity
to verify our intuitions here.

In summary, therefore, the supply barriers are overall neither
negligible nor notably large; the student demand is diffuse, but still
present enought that we believe modest support will elicit this
in a more focused form; the pedagogical pressure is still rather vague
(in the sense that we as teachers are unsure how best to exploit the
resource).  Though these observations fail to be positively
encouraging, they do not undermine our intuition that a relatively
modest technological intervention can have a pronounced and useful --
possibly even transformative -- effect.

\section{The Audiotag system}
\label{s:audiotag}

At the heart of our experiment here is a prototype system, `Audiotag',
developed by one of the authors, which supports upload of audio
recordings, distribution of recordings via podcasts, and collaborative
user tagging of instants within the audio. The system is currently
online on-campus, and the code is available
at \url{https://bitbucket.org/nxg/audiotag/}, under an open licence.
We used versions 0.5 and 0.6 during the course of the session.

Some of the authors have used an earlier version of this system in
previous years, to make recordings available to students in astronomy,
but without laying much stress on the tool, or on the tagging
functionality it offers.

The system
(i) organises and distributes related recordings into `podcasts';
(ii) supports per-use `tagging' of instants within the audio, in a
  manner similar to well-known social websites such as Delicious or
  Flickr;
(iii) supports `likes' of tags, therefore supporting student voting on
  successful or insightful tagging actions; and
(iv) is designed to be coupled to other tools (we are wrestling with
  the pedagogic and user-interface challenges of live tagging via
  mobile devices, in lectures), so that we can support an `ecology' of
  applications which link to, and are linked from, the tagged audio
  instants.
There is a video demo of a recent (but not completely up-to-date)
version of the system at \url{http://vimeo.com/50070137}.

\begin{figure}
\begin{centering}
\includegraphics{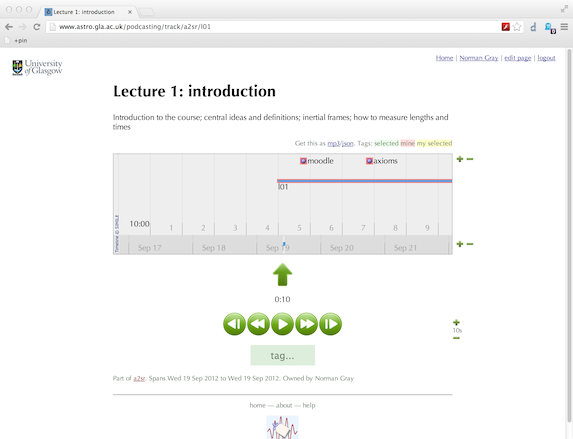}\\
\end{centering}
\caption{\label{f:screenshot}Screenshot of the audiotag web-based application}
\end{figure}

In \prettyref{f:screenshot} we show the user interface to one of the
recordings, showing a recording starting at 10:04 on 19 September
2012, and showing two instants within the opening few minutes tagged
with, respectively `moodle' and `axioms'; this panel can be scrolled
to left and right, and zoomed in and out to show more or less of the
recording.  The user can play, skip and rewind the audio using the
buttons below the display, and add tags to the `current instant' using
the tag box at the bottom.  As well, students can `like' a tag.  The
system is integrated with the university-wide IT identity system, so
that users do not have to register separately for the system.

As well as making recordings available to listen through this
interface, the system also generates a podcast feed so that users can
subscribe to notifications when new recordings are added to a course.

The system has a very simple permissions model: each course has an
`owner', who is typically the lecturer; only the `owner' can upload
recordings, and only logged-in users can add tags, but we have not so
far felt it necessary to restrict access to the audio, so that anyone
can download the lecture audio, and view all the tags, without
authenticating.

\section{Delivering lectures to students -- our experimental evidence this year}
\label{s:experiment}

Two of the authors (NG and NL) have previously used early versions of
the Audiotag server to deliver lecture audio to students, in both
second year and honours, but without laying much stress on it.
Anecdotal evidence suggests that students occasionally used lecture
recordings to catch up on lectures they had missed, but most use was
at revision time, at the end of the session, when students would
listen to complete lectures rather than dropping in to particular
instants; several students reported listening to the lectures whilst
commuting.  There was very little tagging activity in these earlier
presentations, but students spontaneously expressed enthusiasm, both
informally and in course-monitoring questionnaires, for the idea of
making the lectures available.

In session 2012--13 we obtained money, from an
internal Glasgow University learning development fund, to improve
the user interface and to experiment with different
ways of integrating the Audiotag server with other pedagogical
techniques.
Our hope was that we could use the broad insights of the Jigsaw
technique (namely its principled approach to multi-modal group work)
to help students enrich their learning by creating links between their
own lecture notes, pre-distributed lecture notes, and the audio
recordings.

First, however, there is a bootstrap problem.  Before we can create
any dense and multi-modal network of links to tagged audio, we have to
have that tagged audio.  Our experience of previous years suggested
that this was unlikely to happen spontaneously (even though we
believed that we had significantly improved the interface), so we
resorted to an apparently reliable alternative: bribery.  Part of the
financial support was intended as `incentives', which in this
case took the form of Google Nexus~7 tablet computers as prizes for
three of the courses.  We studied six one-semester courses, each of
which was a coherent block of 10 lectures given by a single lecturer,
within a larger full-session course. The collection of courses is
indicated in \prettyref{f:courses}.

\begin{figure}
\begin{tabular}{|rlcccc|}
\hline
\emph{Code}
   & \emph{Course}
   & $N$
   & \emph{Sem}
   & \emph{Year}
   & \emph{Prize?}\\
\hline
a1cos
   & Astronomy 1: Cosmology
   & 112
   & 2nd
   & 1
   & no \\
sats
   & Astronomy 2: Stars and their Spectra
   & 69
   & 2nd
   & 2
   & no \\
cos
   & Honours Astronomy: Cosmology
   & 58
   & 1st
   & honours
   & no \\
e1lds1
   & Exploring the Cosmos
   & 264
   & 2nd
   & 1
   & yes \\
a2sr
   & Astronomy 2: Special Relativity
   & 69
   & 1st
   & 2
   & yes \\
grg1
   & Honours Astronomy: General Relativity
   & 38
   & 1st
   & honours
   & yes \\
\hline
\end{tabular}
\caption{\label{f:courses}The courses studies.  $N$ is the number of
  students in the class; `sem' is the semester in which the class was
  studied (of two); 'year' is the year of study of the students in the
  class, where `honours' represents a mixture of third, fourth and
  fifth-year students; and `prize?' indicates whether one of the
  discussed incentives was available for students.}
\end{figure}

Courses `a1cos', `e1lds1' and `cos' were taught by NL, courses `a2sr'
and grg1 by NG, and `sats' by another colleague in
astronomy\footnote{We are grateful to Matt Pitkin for his willingness
  to experiment here.}.  There
were five other courses this year where lecturers experimented with
the system, and uploaded either a complete or partial set of lectures;
in none were the results obviously different from the three `no-prize'
courses listed above.

These courses represent a broad range of students.  `Exploring the
Cosmos' is a large first-year course often chosen as a filler; while
the students generally enjoy it and are challenged by it (sometimes
more than they expected, under both headings), it is not an academic
priority for many of its students.  `Astronomy 1' and `Astronomy 2'
are required courses for students aiming for astronomy degrees. The
honours courses are both regarded as quite challenging, and are
compulsory or optional for different subsets of the honours class; by
this stage the honours students are highly motivated and in good
command of their learning strategies.

In the three `prize' courses, the class was introduced to the system
via an in-lecture demonstration or pointer to the vimeo.com video
mentioned above, and told that there was a prize -- the tablet
computer -- to be awarded for the `best tagger'; after discussion with
the class, it was decided that this prize would be awarded to the
students whose tags had accumulated the most `likes' by the day of the
course's final exam, in May.  In the `cos', `a2sr' and grg1 courses,
the lecturer added a number of demonstration tags (7, 20, 27
respectively) to the first lecture.   In the three `no-prize' courses,
students were introduced to the system, and encouraged once or twice
to use it.  None of the classes were prescribed any activities
specifically involving the tagging system.

\subsection{Results}
\label{s:results}
From examining the server logs, we discover the RSS (podcast) feeds
for the studied courses were all downloaded on numerous occasions (see
figure below); a single subscription would account for numerous
downloads.  Unfortunately, the server logging available in this
version does not allow us to determine how many unique subscribers
there were or what the RSS clients were, and all we can say at this
point was that we suspect there was only a single subscriber to the
`sats', `cos' and `e1lds1' feeds, or perhaps two
(so at most a few percent of the respective classes),
but that a substantial fraction of
the students in the other courses did subscribe to the podcast feeds.

However many students subscribed to the podcasts, only a very small
number of students have gone on to add tags.  In \prettyref{f:likes}, we
list the number of students who added tags, the number of tags that
they added, and the number of subsequent tag `likes'.
\begin{figure}
\begin{tabular}{|lclcc|}
\hline
\emph{Student}
   & \emph{Course}
   & \emph{Tags (in lectures 1-10)}
   & \emph{Total}
   & \emph{Likes} \\
\hline
KM
   & e1lds1
   & 4, 5, 5, 6, 5, 3, 4, 6, 0, 0 
   & 38
   & 28, 27 \\
HP
   & e1lds1
   & 0, 2, 0, 0, 0, 0, 0, 0, 0, 0
   & 2
   & 1, 1 \\
GA
   & a2sr
   & 0, 9, 0, 0, 28, 16, 0, 0, 0, 0
   & 38
   & \\
KE
   & a2sr
   & 0, 0, 20, 24, 0, 1, 25, 0, 25, 32
   & 127
   & \\
SL
   & a2sr
   & 0, 0, 0, 0, 0, 0, 0, 0, 0, 2
   & 2
   & \\
MG
   & grg1
   & 0, 0, 0, 0, 0, 0, 0, 0, 21, 15
   & 36
   & 2\\
MS
   & grg1
   & 0, 33, 2, 41, 43, 34, 0, 40, 25, 15
   & 233
   & \\
\hline
\end{tabular}
\caption{\label{f:likes}The students who used the `tagging feature.
  `Student' is an (anonymised) label for the student; `course' is one
  of the courses mentioned in \prettyref{f:courses}; `tags' is the
  number of tags added by this student, per lecture, which gives a total
  number in column `total'; and `likes' is the number of times this
  student's tags were `liked' by one or two other students.}
\end{figure}

The three students who tagged extensively (KM, KE and MS) did so
fairly consistently, and the two students who `liked' most, added no
tags themselves.  The students appear to  have added tags fairly
promptly after the lectures, with the exception of KE's, MG's and MS's
tags on their respective lectures 9 and 10, which were tagged
respectively one, one, and four months after the corresponding
lectures.

There are no obvious patterns amongst the students who added tags (and
certainly no patterns significant with so few students).  There is
perhaps a slight overrepresentation of non-native-english speakers --
this \emph{might} be a function of prior educational styles, or of language difficulties.

Our original plan was to use the three first-semester courses to
establish a baseline upon which to investigate the effect of other
pedagogical interventions in semester two.  The surprisingly low
response, however, caused us to change our plans, and make the same
low-intervention observations again to try to establish a more robust
baseline, or to investigate whether there was any difference between
the first and second semesters.

\section{Discussion}
\label{s:summary}
As we discussed in \prettyref{s:background}, we were initially
confident that a technically modest intervention would produce a
significant effect.  This confidence seems to have been misplaced:
either the barriers are higher than we expected, or our intervention
was more modest than is required.  We list some tentative explanations below.

\paraheader{Interface -- general:} User interface design is always harder
than it appears, and it may be that the interface is simply too hard
for users to grasp readily.  We think this is rather unlikely,
however, since the interface has been considerably simplified from
earlier versions of the system, and the informal feedback we have
obtained from students has included suggestions for adjustments
without giving any impression that there is a major usability problem.

\paraheader{Interface -- interaction model:} The implicit interaction
model, in the current design, is that a student will either review a
lecture shortly after it is delivered, or else return to a lecture at
revision time, and work through it adding tags.  While this deliberate
review technique is often suggested to students, we suspect rather few
follow it in fact.  It may be that this interaction model is more
firmly locked in to the system's current interface than we had
thought, so that rather few students are prompted to use it as part
of their existing study habits.  If so, dealing with it would
require either a change in the underlying interaction model, or else
the introduction of explicit exercises to force the students into
interaction.

Over the course of the year, an undergraduate Computing Science
student worked on an alternative interaction model, in which
students use a mobile device to add tags during a
lecture, selecting from a pre-set repertiore of tags\footnote{We thank
  Melissa Campbell for her contributions to the project}.
These tags might represent key 
moments marking `I'm lost here' or `exam', and because they are added
while the user is already interacting with the lecture audio (as live
speech rather than as a recording), they might evade the model-related
problems described above.  Tags such as `I'm lost' are probably most
comfortably kept at least semi-private; this requires a non-trivial
server change, and so while this approach is promising, it was not
possible to fully develop it in this prototype cycle.

One way to align the system's model and the students' is, as above, to
change the system.  An alternative is to change the students: we have
designs for specific exercises which (for example) require the
students to make explicit the links between course handouts and
lecture audio, so forcing an increase in the number of tags, and
thereby intended to create enough value in the set of tags, that
students will interact with the tags completely enough that they cross
a threshold to spontaneously adding more.

\paraheader{Unfamiliarity:} We have supposed that students would be
sufficiently familiar with the concept of tagging online content,
through their experience of existing `Web 2.0' services, that tagging
audio would require no introduction, little training and only mild
encouragement.  It is not obvious that this is false, but until we
have ruled it out, we must consider the possibility that we simply did
not introduce the system clearly enough, so that the students failed
to understand what to do.  If so, this would be a depressingly simple
explanation for the lack of engagement.

\paraheader{Incentive:} The incentive we used on this occasion was a
deliberately generous prize.  Although the nature of an incentive can
sometimes have paradoxical effects on the response, the results above
indicate that the courses where there was tagging activity are
precisely the courses where a prize was offered, so the prize does
seem to have had its intended effect (albeit less pronounced than we
expected).

Overall, this project was a technical success but so far puzzlingly
disappointing in its outcomes.  We initially believed we had rather
small barriers to overcome, namely the barriers dividing students' current practice and
interest from the benefits latent in an easily-obtainable audio
resource.  We expected that we would see rather natural use of
the tagging facilities in the various student populations, so that we
could promptly go on to investigate how this use was changed by
pedagogically motivated exercises.  The results of our investigation
suggest one or more of the following:
(i) that the barriers are higher than we have described in \prettyref{s:background},
or (ii) that we have a poor model of how audio tagging fits
in to students' current practice, or else
(iii, which is not a completely separate issue) that the `natural'
baseline level and pattern of tagging (that is, without forcing from
lecturers' exercises) is significantly lower than the idea of the
`digital native' student might suggest.

In the coming session we plan to repeat the experiment with a modified
interface and a clearer notion of the place of lecturer-driven
exercises, in order to better investigate the shape of the barriers
between students and the latent value of lecture audio recordings.

\bibliography{podcasting}

\begin{thebibliography}{1}

\bibitem{gardiner11}
Kerr Gardiner and Sarah Honeychurch.
\newblock First year student use of technology and their expectations of
  technology use in their courses.
\newblock Technical report, University of Glasgow, 2011.
\newblock URL:
  \url{http://www.gla.ac.uk/services/senateoffice/qea/studentfeedback/thestudentvoice/digitalnatives/}
  [cited May 2013].

\bibitem{aronson13}
Eliot Aronson.
\newblock Jigsaw classroom [online].
\newblock 2013.
\newblock URL: \url{http://www.jigsaw.org} [cited May 2013].

\end{thebibliography}
\bibliographystyle{unsrturl}

\end{document}